\documentclass[letterpaper,12pt,notoc]{JHEP3}

\def\mc{\mathcal}

\usepackage{graphicx}
\usepackage{amsmath}
\usepackage{amssymb}

\preprint{ \hbox{}\hfill arXiv: 1207.1027}

\title{Domain walls in three dimensional gauged
supergravity}
\author{Parinya Karndumri\\
String Theory and Supergravity Group, Department
of Physics, Faculty of Science, Chulalongkorn University, 254 Phayathai Road, Pathumwan, Bangkok 10330, Thailand\\
Thailand Center of Excellence in Physics, CHE, Ministry of Education, Bangkok 10400, Thailand \\
E-mail: \email{parinya.ka@hotmail.com}}

\abstract{We explicitly construct two Chern-Simons gauged
supergravities in three dimensions with $N=4$ and $N=8$
supersymmetries and non-semisimple gauge groups. The $N=4$ theory
has scalar manifold $SO(4,3)/SO(4)\times SO(3)$ with the gauge group
$SO(3)\ltimes (\mathbf{T}^3,\hat{\mathbf{T}}^3)$. The theory
describes $(1,0)$ six dimensional supergravity reduced on an $SU(2)$
group manifold. The equivalent Yang-Mills type gauged supergravity
has $SO(3)$ gauge group coupled to three massive vector fields. The
$N=8$ theory is described by $SO(8,8)/SO(8)\times SO(8)$ scalar
manifold, and the gauge group is given by $SO(8)\ltimes
\mathbf{T}^{28}$. The theory is a truncation of the $SO(8)\ltimes
\mathbf{T}^{28}$ gauged $N=16$ theory with scalar manifold
$E_{8(8)}/SO(16)$ and can be obtained by an $S^7$ compactification
of type I theory in ten dimensions. Domain wall solutions of both
gauged supergravities are analytically found and can be uplifted to
higher dimensions. These provide domain wall vacua in the three
dimensional gauged supergravity framework which might be useful for
the study of Domain Wall$_3$/QFT$_2$ correspondence.}

\begin{document}
\section{Introduction}
Since the original proposal of the AdS/CFT correspondence
\cite{maldacena}, there have been a lot of works generalizing this
conjecture to a more general case than the correspondence between
string theory, or its effective supergravity theory, on an AdS space
and a conformal field theory living on the boundary of the AdS
space. One of the generalizations is the Domain Wall/Quantum Field
Theory (DW/QFT) correspondence \cite{DW/QFT_townsend}. Unlike the
maximally supersymmetric AdS spaces, the domain walls are half
supersymmetric and have only Poincare symmetry in one dimension
lower rather than the full conformal symmetry in the AdS cases. The
solutions are supposed to be dual to the QFT in one dimension lower.
Similar to the standard AdS/CFT correspondence, the correlation
functions of the dual QFT have been studied for example in
\cite{correlator_DW/QFT} and \cite{Skenderis_DW/QFT}. The
correspondence also has an application in cosmology in the form of
domain wall/cosmology correspondence \cite{DW_cosmology},
\cite{DW_cosmology1}, \cite{DW_cosmology2}.
\\
\indent In string/M theory, domain walls arise from the near horizon
limit of p-branes with various dimensions except for D3, M2 and M5
branes for which the dilaton is constant or absent. The resulting
backgrounds are products of the $p+2$ dimensional domain wall and
sphere. In the dual frame, the corresponding metrics can be written
as a product of AdS spaces and spheres with the exception of the
five brane \cite{DW/QFT_townsend}. In supergravities which are
considered as effective theories of string/M theory, they correspond
to domain wall solutions with codimension one of $p+2$ dimensional
gauged supergravity with the gauge group given by the isometry of
the sphere on which ten or eleven dimensional theories are
compactified.
\\
\indent Domain walls in gauged supergravities in various dimensions
have been considered in many places some of which are
\cite{DW_Hull1}, \cite{DW_Hull2}, \cite{Pope_dilatonic_brane},
\cite{Eric_DW1} and \cite{Eric_DW2}. In this work, we focus on the
case of three dimensional gauged supergravity of Chern-Simons type.
The resulting domain walls should correspond to some QFT in two
dimensions. Since gauged supergravity in three dimensions has been
systematically explored only in the last decade, not many studies of
the associated domain wall solutions have been given so far. To the
best of the author's knowledge, the explicit solutions given here
are the first example of the domain wall vacua constructed directly
in this framework. We hope this work will at least provide a toy
model to the study of DW$_3$/QFT$_2$ which, on the other hand, may
also give some insights about the correspondence in higher
dimensions.
\\
\indent Chern-Simons gauged supergravity constructed in
\cite{dewit}, for an earlier construction see \cite{nicolai1} and
\cite{N8}, is the convenient framework to begin with because its
structure is much simpler than the Yang-Mills type theory. A special
class of domain walls in this gauged supergravity in which the
solutions approach AdS spaces in some limits, or equivalently
interpolate between two AdS critical points of the scalar potential,
have been studied in previous works, see for example \cite{bs},
\cite{gkn}, \cite{AP} and \cite{N6}. The resulting solutions have an
interpretation of RG flows describing deformations of a UV CFT to
another CFT in the IR. In this paper, we will study the solution of
the gauged supergravity whose scalar potential does not admit any
critical points on the chosen scalar submanifold. This is the
analogue of the similar study in higher dimensions mentioned above.
\\
\indent We study two gauged supergravities namely $N=4$ and $N=8$
theories with non-semisimple gauge groups. In the $N=4$ case, the
theory describes $(1,0)$ supergravity in six dimensions reduced on
an $SU(2)$ group manifold of $S^3$ topology studied in
\cite{PopeSU2} and \cite{PopeSU22}. The corresponding scalar
manifold is given by $SO(4,3)/SO(4)\times SO(3)$ with the gauge
group being $SO(3)\ltimes (\mathbf{T}^3,\hat{\mathbf{T}}^3)$. This
is equivalent to Yang-Mills type gauged supergravity with gauge
group $SO(3)\sim SU(2)$ and scalar manifold $GL(3,\mathbb{R})/SO(3)$
coupled to three massive vector fields \cite{csym}. The $N=8$ theory
has $SO(8,8)/SO(8)\times SO(8)$ as the scalar manifold in which the
$SO(8)\ltimes \mathbf{T}^{28}\subset SO(8,8)$ is gauged. The theory
is equivalent to $SO(8)$ gauged supergravity of Yang-Mills type and
can be obtained from $S^7$ reduction of type I string theory
\cite{csym}, \cite{Pope_sphere}. The reduced Yang-Mills type theory
does not admit an AdS$_3$ solution but is expected to have a half
supersymmetric domain wall solution which we will explicitly
construct in this paper. The KK spectrum of this reduction has been
given in \cite{Henning_AdS3S7}. Therefore, the domain wall solutions
in both theories can be uplifted to higher dimensions and have
higher dimensional interpretations. The solution in the $N=4$ theory
can be uplifted to a disjoint interior branch of a negative-mass
self-dual string \cite{PopeSU2} while the solution in the $N=8$ case
should be interpreted as a near horizon limit of D1-brane in type I
theory. Furthermore, the theory can be thought of as a truncation of
the $SO(8)\ltimes \mathbf{T}^{28}$ maximal gauged supergravity with
scalar manifold $E_{8(8)}/SO(16)$ arising from $S^7$ reduction of
type IIA or IIB theories in ten dimensions \cite{Henning_AdS3S7},
\cite{nicolai3}.
\\
\indent The paper is organized as follow. We give a short discussion
on the structure of three dimensional gauged supergravity in section
\ref{3DSUGRA}. We refer the reader to \cite{dewit} for the full
discussion. We then construct $N=4$ gauged supergravity with scalar
manifold $SO(4,3)/SO(4)\times SO(3)$ and gauge group $SO(3)\ltimes
(\mathbf{T}^3,\hat{\mathbf{T}}^3)$ in section \ref{N4theory}. The
associated domain wall solution is also given in this section. In
section \ref{N8theory}, we move on to the $N=8$ theory in which the
scalar manifold is $SO(8,8)/SO(8)\times SO(8)$, and the gauge group
is given by $SO(8)\ltimes \mathbf{T}^{28}$. We also find a domain
wall solution in this case. The appendix is devoted to some useful
formulae for the $N=8$ theory. Apart from giving more details, the
aim of this appendix is also to give the more detailed construction
for further investigations of the resulting scalar potential which
is useful in the study of holographic RG flows. We end the paper
with some conclusions and comments in section \ref{conclusion}. All
computations carried out in this work are achieved by using the
computer program \textsl{Mathematica}.

\section{Gauged supergravity in three dimensions}\label{3DSUGRA}
We now give a brief review of three dimensional gauged supergravity
in the $SO(N)$ R-symmetry covariant formulation of \cite{dewit}. We
refer the reader to \cite{dewit} for both details and notations. The
gauged supergravity theory is described by gaugings of non-linear
sigma models coupled to supergravity which is by itself topological.
In this work, we always consider the case of symmetric scalar target
spaces of the form $G/H$. Specifically, we are interested in the
$N=4$ and $N=8$ theories with scalar manifolds $SO(4,3)/SO(4)\times
SO(3)$ and $SO(8,8)/SO(8,8)$, respectively.
\\
\indent Couplings of the sigma model to supergravity require $N-1$
almost complex structures $f^P$, $P=2,\ldots, N$ which can be used
to find the tensors $f^{IJ}$ as follow:
\begin{equation}
f^{1P}=-f^{P1}=f^P,\qquad f^{PQ}=f^{[P}f^{Q]}\, .\label{fIJ}
\end{equation}
They generate the $SO(N)$ R-symmetry in the spinor representation.
As in \cite{dewit}, indices $I,J=1,\ldots N$ are R-symmetry indices
while $i,j=1,\ldots, d$, or in the flat basis $A,B=1,\ldots, d$,
label coordinates on the target space of dimension $d$.
\\
\indent In general, scalar manifolds of the $N=4$ theory are product
of two quaternionic manifolds which need not be symmetric. The
R-symmetry $SO(4)\sim SO(3)\times SO(3)$ acts separately on the two
subspaces. We will study the case in which the target space consists
of only one quaternionic manifold, sometime called the degenerate
case. On the other hand, for the $N=8$ theory, supersymmetry
requires that the scalar manifolds are of the form
$SO(8,k)/SO(8)\times SO(k)$ \cite{dewit1}. $k$ labels the number of
matter supermultiplets. Before moving on, we note some useful
formulae for the coset spaces. The coset space $G/H$ can be
described by the coset representative $L$ transforming by left and
right multiplications under the global $G$ and local $H$ symmetries.
With $t^{\mc{M}}$, $\mc{M}=1,\ldots, \textrm{dim}G$, denoting $G$
generators which can be decomposed into $H=SO(N)\times H'$
generators and the non-compact ones as $\{T^{IJ},X^\alpha,Y^A\}$,
for $\alpha=1,\ldots, \textrm{dim}H'$ and $A=1,\ldots ,
d(\textrm{dim}(G/H))$, we have formulae
\begin{eqnarray}
L^{-1}t^\mathcal{M}L&=&\frac{1}{2}\mathcal{V}^{\mathcal{M}}_{\phantom{as}IJ}T^{IJ}+\mathcal{V}^\mathcal{M}_{\phantom{as}\alpha}X^\alpha+
\mathcal{V}^\mathcal{M}_{\phantom{as}A}Y^A,\label{cosetFormula}\\
L^{-1} \partial_i L&=& \frac{1}{2}Q^{IJ}_i T^{IJ}+Q^\alpha_i
X^{\alpha}+e^A_i Y^A\, . \label{cosetFormula1}
\end{eqnarray}
Generators $t^{\mc{M}}$'s satisfy the algebra
\begin{eqnarray}
\left[T^{IJ},T^{KL}\right]&=&-4\delta^{[I[K}T^{L]J]}, \qquad
\left[T^{IJ},Y^A\right]=-\frac{1}{2}f^{IJ,AB}Y_B, \nonumber \\
\left[X^\alpha,X^\beta\right]&=&f^{\alpha
\beta}_{\phantom{as}\gamma}X^\gamma,\qquad
\left[X^\alpha,Y^A\right]=h^{\alpha
\phantom{a}A}_{\phantom{a}B}Y^B, \nonumber \\
\left[Y^{A},Y^{B}\right]&=&\frac{1}{4}f^{AB}_{IJ}T^{IJ}+\frac{1}{8}C_{\alpha\beta}h^{\beta
AB}X^\alpha \label{Galgebra}
\end{eqnarray}
which will be important later on. In the above algebra, the
anti-symmetric tensors $h^{\alpha \phantom{a}A}_{\phantom{a}B}$
generate the $H'$ algebra in the spinor representation of $SO(N)$
with structure constants $f^{\alpha \beta}_{\phantom{as}\gamma}$,
and $C_{\alpha\beta}$ are symmetric $H'$ invariant tensors, see
\cite{dewit1} for more information about this algebra. The
non-compact generators $Y^A$ transform as a spinor representation of
$SO(N)$. The tensors $f^{IJ}$ can be written in terms of $SO(N)$
gamma matrices $\Gamma^I$ as
\begin{equation}
f^{IJ}=-\frac{1}{2}\Gamma^{IJ}=-\frac{1}{4}\left(\Gamma^I\Gamma^J-\Gamma^J\Gamma^I\right).
\end{equation}
\\
\indent Gaugings can be accomplished by introducing the so-called
embedding tensor. In three dimensions, this tensor lives in a
symmetric product of two adjoint representations of $G$. The product
is in turn decomposed into irreducible representations of $G$. For
viable gaugings, the corresponding embedding tensors have to satisfy
two constaints
\begin{eqnarray}
\Theta_{\mathcal{PL}}f^{\mathcal{KL}}_{\phantom{asds}\mathcal{(M}}\Theta_{\mathcal{N)K}}&=&0,\label{theta_quadratic}\\
\mathbb{P}_{R_0}\Theta_{\mc{MN}}&=&0\, .\label{Theta_constraint}
\end{eqnarray}
The first constraint is a requirement for the gauge generators,
$J_\mc{M}=\Theta_{\mc{MN}}t^\mc{N}$, to form a proper subalgebra of
$G$. The second one comes from supersymmetry. In general,
supersymmetry requires the T-tensor, defined below, to satisfy
\begin{equation}
\mathbb{P}_\boxplus T^{IJ,KL}=0\label{Tconstraint1}
\end{equation}
where $\boxplus$ denotes the Riemann tensor-like representation of
$SO(N)$. But, for symmetric target spaces, this constraint can be
uplifted to be the constraint on $\Theta_{\mc{MN}}$ given in
\eqref{Theta_constraint} \cite{dewit}. The $R_0$ is a unique $G$
representation whose branching under $SO(N)$ contains the $\boxplus$
representation of $SO(N)$. In this case, the condition
\eqref{Tconstraint1} implies \eqref{Theta_constraint}.
\\
\indent Gaugings introduce minimal couplings through covariant
derivatives, and the gauge fields enter the gauged Lagrangian via
the Chern-Simons terms. To restore supersymmetry broken by the
gaugings, fermionic mass-like terms and the scalar potential are
needed. These terms are written in terms of the T-tensor given by
\begin{equation}
T_{\mathcal{AB}}=\mathcal{V}^{\mc{M}}_{\phantom{as}\mc{A}}\Theta_{\mc{MN}}\mathcal{V}^{\mc{N}}_{\phantom{as}\mc{B}}\,
.\label{T_tensor_def}
\end{equation}
$\mc{A},\mc{B}$ indices decompose into $\{IJ, \alpha, A\}$ in which
$IJ$ and $\alpha$ label adjoint representations of $SO(N)$ and $H'$,
and $A$ is the $SO(N)$ spinor index. The map
$\mc{V}^{\mc{M}}_{\phantom{as}\mc{A}}$ can be obtained from
\eqref{cosetFormula}. The $A_1$ and $A_2$ tensors are needed in
order to compute the scalar potential as well as the supersymmetry
transformations of fermions. They are given by
\begin{eqnarray}
A_1^{IJ}&=&-\frac{4}{N-2}T^{IM,JM}+\frac{2}{(N-1)(N-2)}\delta^{IJ}T^{MN,MN},\nonumber\\
A_{2j}^{IJ}&=&\frac{2}{N}T^{IJ}_{\phantom{as}j}+\frac{4}{N(N-2)}f^{M(I
m}_{\phantom{as}j}T^{J)M}_{\phantom{as}m}+\frac{2}{N(N-1)(N-2)}\delta^{IJ}f^{KL\phantom{a}m}_{\phantom{as}j}T^{KL}_{\phantom{as}m},
\nonumber \\
V&=&-\frac{4}{N}g^2\left(A_1^{IJ}A_1^{IJ}-\frac{1}{2}Ng^{ij}A_{2i}^{IJ}A_{2j}^{IJ}\right).
\end{eqnarray}
\\
\indent We will also need the supersymmetry transformations of
fermions in order to find supersymmetric solutions. There are $N$
gravitini $\psi^I_\mu$ and $d$ spin $\frac{1}{2}$ fields
$\chi^{iI}$. Notice that $\chi^{iI}$ are written in an overcomplete
basis. They are subject to the projection constraint given in
\cite{dewit} such that the total number of independent $\chi^{iI}$
is $d$. Indices $\mu,\nu=0,1,2$ denote three dimensional spacetime
coordinates. The supersymmetry transformations are then given by
\cite{dewit}
\begin{eqnarray}
\delta\psi^I_\mu
&=&\mathcal{D}_\mu\epsilon^I+gA_1^{IJ}\gamma_\mu\epsilon^J,\nonumber\\
\delta\chi^{iI}&=&
\frac{1}{2}(\delta^{IJ}\mathbf{1}-f^{IJ})^i_{\phantom{a}j}{\mathcal{D}{\!\!\!\!/}}\phi^j\epsilon^J
-gNA_2^{JIi}\epsilon^J\label{susyvar}
\end{eqnarray}
where we have set all the fermionic fields to zero on the right hand
side. The covariant derivative $\mathcal{D}_\mu$ includes the usual
spin connection. In this paper, we will use the metric signature
$(-++)$ rather than the Pauli-Kallen metric used in \cite{dewit},
see also the transformation to $(-++)$ metric given in
\cite{dewit1}.
\\
\indent We now discuss the relevant form of the embedding tensor
used throughout this paper. The gauge group is a non-semisimple
group of the form $G_0\ltimes
(\mathbf{T}^{\textrm{dim}G_0},\hat{\mathbf{T}}^M)\subset G$.
$\mathbf{T}^{\textrm{dim}G_0}$ consists of $\textrm{dim}G_0$
commuting generators transforming in the adjoint representation of
$G_0$ while $M$ nilpotent generators of $\hat{\mathbf{T}}^M$
transform in some $M$ dimensional representation of $G_0$ and close
onto the translational generators $\mathbf{T}^{\textrm{dim}G_0}$. It
has been shown in \cite{csym} that the corresponding embedding
tensor is given by
\begin{equation}
\Theta
=g_1\Theta_{\mc{A}\mc{B}}+g_2\Theta_{\mc{B}\mc{B}}+g_3\Theta_{\mc{H}\mc{H}}
\end{equation}
where $\mc{A}$, $\mc{B}$ and $\mc{H}$ refer to $G_0$,
$\mathbf{T}^{\textrm{dim}G_0}$ and $\hat{\mathbf{T}}^M$ parts of the
full gauge group, respectively. Although the indices $\mc{A}$ and
$\mc{B}$ have been used previously as indices of the T-tensors,
there should be no confusions here since the latter no longer appear
in the rest of the paper. This gauging is on-shell equivalent to
Yang-Mills gauged supergravity with gauge group $G_0$ coupled to $M$
massive vector fields \cite{csym}. Notice that there is no coupling
between the semisimple part with itself. This is a crucial point for
the CS-YM equivalence to work \cite{csym}. For later convenience, we
will also repeat the full gauge algebra from \cite{csym}
\begin{eqnarray}
\left[J^m,J^n\right]&=&f^{mn}_{\phantom{ass}k}J^k,\qquad
\left[J^m,T^n\right]=f^{mn}_{\phantom{ass}k}T^k,\qquad
\left[T^m,T^n\right]=0,\nonumber \\
\left[J^m,\hat{T}^\alpha\right]&=&t^{m\alpha}_{\phantom{sas}\beta}\hat{T}^\beta,\qquad
\left[\hat{T}^\alpha,\hat{T}^\beta\right]=t^{\alpha\beta}_{\phantom{ass}m}T^m,\qquad
\left[T^m,\hat{T}^\alpha\right]=0,\label{G0_algebra}
\end{eqnarray}
where $J^m$, $T^m$ and $\hat{T}^\alpha$ are generators of $G_0$,
$\mathbf{T}^{\textrm{dim}G_0}$ and $\hat{\mathbf{T}}^M$,
respectively. The $\Theta_{\mc{AB}}$ and $\Theta_{\mc{BB}}$ are
given by the Cartan-Killing form, $\eta_{mn}$, $m,n=1,\ldots,
\textrm{dim}G_0$, while the $\Theta_{\mc{HH}}$ is described by
$\kappa_{\alpha\beta}$ given by $\eta_{mn}$ and the structure
constants in \eqref{G0_algebra} via the relation
$\eta_{mn}t^{n\alpha}_{\phantom{asd}\beta}=\kappa_{\beta\gamma}t^{\alpha\gamma}_{\phantom{sad}m}$
\cite{csym}. The couplings $g_1$, $g_2$ and $g_3$ are in general not
independent. As we will see, consistency conditions will impose some
relations between them.

\section{A domain wall solution in $N=4$ theory}\label{N4theory}
In this section, we study a domain wall solution in the $N=4$ gauged
supergravity with $SO(4,3)/SO(4)\times SO(3)$ scalar manifold. The
gauge group considered here is the non-semisimple group
$SO(3)\ltimes (\mathbf{T}^3,\hat{\mathbf{T}}^3)$. As mentioned
before, this theory describes the reduction of $N=(1,0)$
supergravity in six dimensions on an $SU(2)$ group manifold. The
reduced theory is equivalent to the $SO(3)\ltimes
(\mathbf{T}^3,\hat{\mathbf{T}}^3)$ CS type gauged supergravity in
three dimensions.
\\
\indent We begin with the structure of the coset manifold. We use
the basis elements of the general $7\times 7$ matrices
\begin{equation}
(e_{ab})_{cd}=\delta_{ac}\delta_{bd},\qquad a,b=1,\ldots , 7\, .
\end{equation}
These are generators of $GL(7,\mathbb{R})$. There are 12 scalars in
the coset $SO(4,3)/SO(4)\times SO(3)$. Under the maximal compact
subgroup $SO(4)\times SO(3)\sim SO(3)\times SO(3)\times SO(3)$, they
transform as
\begin{equation}
(\mathbf{4},\mathbf{3})=(\mathbf{2},\mathbf{2},\mathbf{3}).
\end{equation}
We can see that the scalars transform as a spinor representation
under the first two $SO(3)$'s in the $SO(4)$. We then identify one
of them as the R-symmetry group $SO(3)_R$. This choice is certainly
not unique. The crucial point in this formulation is the fact that
the non-compact generators transform as a spinor of $SO(3)$. We can
also choose the last $SO(3)$ to be the R-symmetry provided that we
change the basis in such a way that the non-compact generators
transform as a spinor representation under this $SO(3)$.
\\
\indent To label the three different $SO(3)$'s, we use the following
identification: $SO(3)_R\times SO(3)'\subset SO(4)$ and
$SO(3)^{(2)}$ for the $SO(3)$ in the maximal compact subgroup of
$SO(4,3)$. We now construct generators of $SO(4,3)$. The
$SO(4)\times SO(3)$ generators are given by
\begin{eqnarray}
SO(3)^{(2)}&:&\qquad J_1^{ij}=e_{ji}-e_{ij},\qquad i,j=1,2,3,\nonumber \\
SO(4)&:&\qquad j^{ab}=e_{b+3,a+3}-e_{a+3,b+3},\qquad a,b=1,\ldots,
4\, .
\end{eqnarray}
The corresponding generators of $SO(3)_R\times SO(3)'$ are then
given by
\begin{eqnarray}
SO(3)_R &:& \bar{T}^1=j^{12}+j^{34},\qquad \bar{T}^2=j^{13}-j^{24},\qquad \bar{T}^3=j^{23}+j^{14}, \nonumber \\
SO(3)' &:& \tilde{T}=j^{12}-j^{34},\qquad
\tilde{T}^2=j^{13}+j^{24},\qquad \tilde{T}^3=j^{23}-j^{14}\, .
\end{eqnarray}
However, it is more convenient in the computation to label $SO(3)_R$
generators by $T^{IJ}$ with $I,J=1,2,3,4$. Accordingly, we define
the following generators
\begin{equation}
SO(3)_R :\qquad
T^{IJ}=\frac{1}{2}\left(j^{IJ}+\frac{1}{2}\epsilon_{IJKL}j^{KL}\right),\qquad
I,J=1,\ldots, 4\, .
\end{equation}
Finally, the non-compact generators are given by
\begin{equation}
Y^A=\left\{
      \begin{array}{ll}
        \frac{1}{\sqrt{2}}(e_{1,A+3}+e_{A+3,1}), & A=1,\ldots, 4, \\
        \frac{1}{\sqrt{2}}(e_{2,A-1}+e_{A-1,2}), & A=5,\ldots, 8,\\
        \frac{1}{\sqrt{2}}(e_{3,A-5}+e_{A-5,3}), & A=9,\ldots, 12\,
        .
      \end{array}
    \right.
\end{equation}
We have labeled the non-compact generators such that they fit into
our general formulation. Furthermore, the normalization of $T^{IJ}$
and $Y^A$ is chosen to satisfy the G-algebra \eqref{Galgebra}.
\\
\indent We now move to the generators of the gauge group. First of
all, the semisimple part of the gauge group $SO(3)$ is given by the
diagonal of all three $SO(3)$'s, $SO(3)=\left[SO(3)_R\times
SO(3)'\times SO(3)^{(2)}\right]_{\textrm{diag}}$. The corresponding
generators are given by
\begin{equation}
J^{ij}=J_1^{ij}+j^{i+1,j+1},\qquad i,j=1,2,3\, .
\end{equation}
The translational generators are given by
\begin{equation}
t^{ij}=J_1^{ij}-j^{ij}-(e_{j+4,i}+e_{i,j+4})+(e_{j,i+4}+e_{i+4,j}),\qquad
i,j=1,2,3\, .
\end{equation}
It can be verified that they transform as an adjoint representation
of $SO(3)$ and that they commute with each other. Finally, the
nilpotent generators of $\hat{\mathbf{T}}^3$ are found to be
\begin{equation}
\hat{t}^\alpha=e_{\alpha4}+e_{4\alpha}+e_{4,\alpha+4}-e_{\alpha+4,4},\qquad
\alpha=1,2,3,
\end{equation}
which transform as a vector representation of $SO(3)$ and close onto
$\mathbf{T}^3$.
\\
\indent Recall that the general form of the embedding tensor for
this gauge group is given by
\begin{equation}
\Theta
=g_1\Theta_{\mc{AB}}+g_2\Theta_{\mc{BB}}+g_3\Theta_{\mc{H}\mc{H}}\,
.
\end{equation}
The symbols $\mc{A}$, $\mc{B}$ and $\mc{H}$ denote the $SO(3)$,
$\mathbf{T}^3$ and $\hat{\mathbf{T}}^3$ parts of the full gauge
group, respectively. It turns out that consistency conditions
\eqref{theta_quadratic} and \eqref{Theta_constraint} require
\begin{equation}
g_2=0,\qquad g_3=-g_1=-g\, .
\end{equation}
This implies that the $SO(3)\times \mathbf{T}^3$ is by itself not a
consistent gauge group. If this gauge group was possible, the
corresponding Yang-Mills type would describe the $SU(2)$ reduction
of the $(1,0)$ six dimensional theory without any massive vector
fields. This is consistent with the fact that the massive vector
fields cannot be truncated out without truncating the gauge fields
or setting the two-form flux along $S^3$ to zero. This is because,
from the reduced equation of motion for the massive vector fields
given in \cite{PopeSU2}, the gauge fields act as a source term for
the massive vector fields.
\\
\indent The tensor $f^{IJ}$ can be computed from the G-algebra
\eqref{Galgebra} by
\begin{equation}
f^{IJ}_{AB}=-2\textrm{Tr}(Y_B\left[T^{IJ},Y_A\right]).
\end{equation}
All the $\mc{V}$ maps are calculated by the following relations
coming from \eqref{cosetFormula}
\begin{eqnarray}
\mc{V}_{\mc{A}}^{m,IJ}&=&-\textrm{Tr}(L^{-1}X^mLT^{IJ}),\qquad
\mc{V}_{\mc{B}}^{m,IJ}=-\textrm{Tr}(L^{-1}S^mLT^{IJ}),\qquad m=1,2,3,\nonumber \\
\mc{V}_{\mc{H}}^{m,IJ}&=&-\textrm{Tr}(L^{-1}\hat{t}^mLT^{IJ}),\qquad
\mc{V}_{\mc{A}}^{m,A}=\textrm{Tr}(L^{-1}X^mLY^{A}),\nonumber \\
\mc{V}_{\mc{B}}^{m,A}&=&\textrm{Tr}(L^{-1}S^mLY^{A}),\qquad
\mc{V}_{\mc{H}}^{m,A}=\textrm{Tr}(L^{-1}\hat{t}^mLY^{A})
\end{eqnarray}
where for conveniences, we have redefined the gauge generators to be
\begin{eqnarray}
X^1&=&J^{12},\qquad X^2=J^{13},\qquad X^3=J^{23},\nonumber \\
S^1&=&t^{12},\qquad S^2=t^{13},\qquad S^3=t^{23}\, .
\end{eqnarray}
\indent Under the $SO(3)=\left[SO(3)_R\times SO(3)'\times
SO(3)^{(2)}\right]_{\textrm{diag}}$ of the gauge group, the 12
scalars transform as
\begin{equation}
(\mathbf{2}\times \mathbf{2})\times
\mathbf{3}=\mathbf{1}+\mathbf{3}_V+\mathbf{3}_A+\mathbf{5}
\end{equation}
where we have distinguished the two representation $\mathbf{3}$'s by
the subscripts $V$ (vector) and $A$ (anti-symmetric tensor). This
decomposition gives a new basis for the 12 scalars. The above
representations correspond to the following non-compact generators
\begin{eqnarray}
\mathbf{1}:\qquad
Y^{(1)}&=&\frac{1}{\sqrt{3}}\left(Y_2+Y_7+Y_{12}\right),\nonumber \\
\mathbf{3}_A:\qquad
Y^{(3)_A}_1&=&\frac{1}{\sqrt{2}}\left(Y_3-Y_6\right),\qquad
Y^{(3)_A}_2=\frac{1}{\sqrt{2}}\left(Y_4-Y_{10}\right),\nonumber \\
Y^{(3)_A}_3&=&\frac{1}{\sqrt{2}}\left(Y_8-Y_{11}\right),\nonumber \\
\mathbf{3}_V:\qquad Y^{(3)_V}_1&=&Y_1,\qquad Y^{(3)_V}_2=Y_5,\qquad
Y^{(3)_V}_3=Y_9,\nonumber \\
\mathbf{5}:\qquad Y^{(5)}_1
&=&\frac{1}{\sqrt{2}}\left(Y_3+Y_6\right),\qquad Y^{(5)}_2
=\frac{1}{\sqrt{2}}\left(Y_4+Y_{10}\right),\nonumber \\ Y^{(5)}_3
&=&\frac{1}{\sqrt{2}}\left(Y_8+Y_{11}\right),\qquad
Y^{(5)}_4 =\frac{1}{\sqrt{2}}\left(Y_2-Y_7\right),\nonumber \\
Y^{(5)}_5 &=&\frac{1}{\sqrt{3}}\left(Y_2+Y_7-2Y_{12}\right).
\end{eqnarray}
\indent Given the form of the coset representative $L$, we can now
compute the T-tensor, $A_1$ and $A_2$ tensors and finally the scalar
potential from the embedding tensor via the formulae give in the
previous section. Explicitly, the relevant components of the
T-tensor are given by
\begin{eqnarray}
T^{IJ,KL}&=&g(\mc{V}_{\mc{A}}^{m,IJ}\mc{V}_{\mc{B}}^{m,KL}+\mc{V}_{\mc{B}}^{m,IJ}\mc{V}_{\mc{A}}^{m,KL})\nonumber
\\
& &
-g\mc{V}_{\mc{H}}^{m,IJ}\mc{V}_{\mc{H}}^{m,KL}\\
T^{IJ,A}&=&g(\mc{V}_{\mc{A}}^{m,IJ}\mc{V}_{\mc{B}}^{m,A}+\mc{V}_{\mc{B}}^{m,IJ}\mc{V}_{\mc{A}}^{m,A})\nonumber
\\
& & -g\mc{V}_{\mc{H}}^{m,IJ}\mc{V}_{\mc{H}}^{m,A}\, .
\end{eqnarray}
It turns out to be very complicated to compute the scalar potential
for all 12 scalars simultaneously. We then study the potential in
each sector according to the representations under $SO(3)$ given
above. For scalar in the $\mathbf{5}$, $\mathbf{3}_A$ and
$\mathbf{3}_V$, there are no interesting non-trivial critical
points. Therefore, we will not study these cases further. Moreover,
we will not give the explicit form of the resulting potential since
they are not relevant for this work.
\\
\indent When turning on the $SO(3)$ singlet scalar, we find that the
scalar potential does not have any critical points. Rather than a
maximally supersymmetric $AdS_3$ critical point at $L=\mathbf{I}$,
there is a domain wall solution preserving half of the supersymmetry
as we will see. We begin with the coset representative
\begin{equation}
L=e^{a(r)Y^{(1)}}\, .
\end{equation}
The scalar potential is computed to be
\begin{equation}
V=-96e^{\sqrt{\frac{8}{3}}a(r)}g^2
\end{equation}
which clearly does not have any critical points. We then expect to
find a domain wall solution in this case. Notice that the potential
take the same form as those studied in higher dimensional gauged
supergravities.
\\
\indent We want to find a half-supersymmetric solution, so we begin
with the BPS equations coming from setting the supersymmetry
variations of the fermionic fields $\psi^I_\mu$ and $\chi^{iI}$ to
zero. The metric is given by the usual domain wall ansatz
\begin{equation}
ds^2=e^{2A(r)}dx^2_{1,1}+dr^2\, .\label{DW_ansatz}
\end{equation}
Generally, in the $D$ dimensional domain wall ansatz, we want the
$D-1$ dimensional Poincare symmetry to be preserved. Therefore, the
function $A$ can depend only on the radial coordinate $r$. Any
function in front of $dr^2$ can be absorbed by the redefinition of
$r$. So, the above ansatz is the general ansatz for the domain wall
preserving two dimensional Poincare symmetry $ISO(1,1)$. The ansatz
is similar to that in the study of holographic RG flows in which the
metric interpolates between two $AdS_3$ spaces. In that case, the
function $A(r)$ behaves linearly in $r$ at both ends. It is
well-known that an AdS space is a special case of domain walls for
which the isometry gets enhanced from $ISO(1,1)$ to $SO(2,2)$ which
is the isometry of $AdS_3$. The number of supersymmetry is twice
that of the domain wall as well.
\\
\indent The solutions studied here are domain walls that do not give
$AdS_3$ as a limiting case. We will explicitly see below that the
solution for $A(r)$ will not be linear in any limits of $r$. The
solution will rather be a flat space $\mathbb{R}^{1,2}$ at one limit
in $r$, at $r=0$ or $r=\infty$. We now put the corresponding spin
connection computed from the metric into the BPS equations
$\delta\chi^{iI}=0$ and $\delta\psi^I_\mu=0$ for $\mu=0,1$. With the
condition $\gamma_r\epsilon^I=\epsilon^I$, the former gives
\begin{equation}
a'+2\sqrt{6}e^{\sqrt{\frac{2}{3}}a}g=0\label{a_eq}
\end{equation}
while the latter gives
\begin{equation}
A'-12e^{\sqrt{\frac{2}{3}}a}g=0\, .\label{A_eq}
\end{equation}
Due to the projection by $\gamma_r$, the resulting solution will be
half-supersymmetric. In the above equations and the remaining ones
in the paper, we have used the notation $a'=\frac{da}{dr}$. Equation
\eqref{a_eq} can be easily solved to obtain
\begin{equation}
a(r)=-\sqrt{\frac{3}{2}}\ln \left(8gr-\sqrt{\frac{2}{3}}C_1\right)
\end{equation}
where $C_1$ is a constant. Inserting the solution for $a$ into
\eqref{A_eq}, we can solve for the $A$ solution
\begin{equation}
A(r)=C_2+\frac{3}{2}\ln \left(24gr-\sqrt{6}C_1\right).
\end{equation}
\indent The $\delta\psi^I_r=0$ equation gives the condition on the
$r$-dependent Killing spinors. With the ansatz
$\epsilon^I=e^{f(r)}\epsilon^I_0$ for
$\gamma_r\epsilon^I_0=\epsilon^I_0$, it can be easily verified that
$\delta\psi^I_r=0$ equation is satisfied by
\begin{equation}
\epsilon^I=e^{\frac{A}{2}}\epsilon_0^I
\end{equation}
similar to the corresponding solutions in higher dimensions.
\\
\indent The constant $C_2$ can be set to zero by rescaling the
coordinates $x^0$ and $x^1$. On the other hand, by shifting the
coordinate $r$, we can remove the constant $C_1$. The metric is then
given by
\begin{equation}
ds^2=(24g_1r)^3dx_{1,1}^2+dr^2\, .\label{N4Sol_metric}
\end{equation}
We can also write it in the form of the warped $AdS_3$. To do this,
we first rescale all of the coordinates to the dimensionless ones,
recalling that the coupling $g$ has a dimension of mass,
\begin{equation}
\tilde{x}^0= gx^0,\qquad \tilde{x}^1= gx^1,\qquad \textrm{and}
\qquad \tilde{r}=gr\, .
\end{equation}
The metric \eqref{N4Sol_metric} can then be rewritten as
\begin{equation}
ds^2=\frac{1}{3^{15}(2^{12})g^2}\rho^{-4}\left(\frac{dx^2_{1,1}+d\rho^2}{\rho^2}\right)
\end{equation}
by using the new coordinate
$\rho=-\frac{2}{3(24)^{\frac{3}{2}}}r^{-\frac{1}{2}}$.

\section{A domain wall solution in $N=8$ theory}\label{N8theory}
In this section, we will study $N=8$ gauged supergravity in three
dimensions with scalar target space $SO(8,8)/SO(8)\times SO(8)$.
This theory has also been studied in \cite{gkn} with another gauge
group $(SO(4)\ltimes \mathbf{T}^6)\times (SO(4)\ltimes
\mathbf{T}^6)$. In that case, the theory is supposed to describe the
nine dimensional supergravity reduced on $S^3\times S^3$. Here, we
will study the same theory with gauge group $SO(8)\ltimes
\mathbf{T}^{28}$. According to the equivalence to the $SO(8)$
Yang-Mills gauged supergravity \cite{csym}, this theory should
describe the compactification of the ten dimensional type I theory
on $S^7$.
\\
\indent As in the $N=4$ theory, we still use the general formulation
of \cite{dewit}. We begin with the structure of $SO(8,8)/SO(8)\times
SO(8)$ coset. We will use the coset representative in the
fundamental representation $\mathbf{16}$ of $SO(8,8)$. The 64
scalars transform as a bivector, $(\mathbf{8},\mathbf{8})$, of the
two $SO(8)$'s. However, we need the scalars to transform as
$(\mathbf{8}_s,\mathbf{8})$ in order to fit into the $SO(N)$
covariant formulation of \cite{dewit}. We note here our notations
regarding to the $SO(8)$ representations. The vector representation
of $SO(8)$ is simply denoted by $\mathbf{8}$ while the spinor and
conjugate spinor representations are denoted by $\mathbf{8}_s$ and
$\mathbf{8}_c$, respectively.
\\
\indent Similar to \cite{gkn}, we will use the $SO(8)$ R-symmetry in
the spinor representation of the form
\begin{equation}
T^{IJ}=\left(
         \begin{array}{cc}
           \Gamma^{IJ} & 0 \\
           0 & 0 \\
         \end{array}
       \right)
\end{equation}
where
$\Gamma^{IJ}=-\frac{1}{4}\left(\Gamma^I(\Gamma^J)^T-\Gamma^J(\Gamma^I)^T\right)$.
The $\Gamma^I$ are $8\times 8$ gamma matrices of $SO(8)$ whose
explicit forms are given in the appendix. They are embedded in the
full Dirac gamma matrices $\gamma^I$ as
\begin{equation}
\gamma^I=\left(
           \begin{array}{cc}
             0 & \Gamma^I \\
             (\Gamma^I)^T & 0 \\
           \end{array}
         \right).
\end{equation}
It can be easily verified that $T^{IJ}$ satisfy $SO(8)$ algebra
given in \eqref{Galgebra}. We have used the same notation for
$SO(8)$ Dirac gamma matrices and the spacetimes gamma matrices, but
this should not be confusing since the former will not appear
elsewhere in the paper.
\\
\indent In order to construct the $SO(8,8)/SO(8)\times SO(8)$ coset,
we define the generators of $GL(16,\mathbb{R})$
\begin{equation}
(e_{mn})_{pq}=\delta_{mp}\delta_{nq},\qquad m,n,p,q=1,\ldots,16\, .
\end{equation}
The maximal compact subgroup $SO(8)^{(1)}\times SO(8)^{(2)}$ is
generated by
\begin{eqnarray}
SO(8)^{(1)}&:&\qquad J_1^{ab}=e_{ba}-e_{ab},\qquad a,b =1,\ldots,
8,\nonumber \\
SO(8)^{(2)}&:&\qquad J_2^{rs}=e_{s+8,r+8}-e_{r+8,s+8} ,\qquad r,s
=1,\ldots, 8\, .
\end{eqnarray}
The non-compact generators transforming in the
$(\mathbf{8},\mathbf{8})$ representation of $SO(8)^{(1)}\times
SO(8)^{(2)}$ are given by
\begin{equation}
Y^{ar}=e_{a,r+8}+e_{r+8,a},\qquad r,a =1,\ldots , 8\, .
\end{equation}
\indent We now come to $SO(8)\ltimes \mathbf{T}^{28}$ generators.
The semisimple part is given by the diagonal subgroup of
$SO(8)^{(1)}\times SO(8)^{(2)}$. Including the 28 commuting
generators, $t^{ab}=t^{[ab]}$, of $\mathbf{T}^{28}$, the full gauge
generators are given by
\begin{eqnarray}
SO(8)&:&\qquad J^{ab}=J^{ab}_1+J^{ab}_{2},\nonumber \\
\mathbf{T}^{28} &:& \qquad t^{ab}=J_1^{ab}-J_2^{ab}+Y^{ba}-Y^{ab}\,
.
\end{eqnarray}
We can check that they satisfy the algebra \eqref{G0_algebra}.
Similar to \cite{gkn}, the $f^{IJ}$ tensor is defined by
\begin{equation}
f^{IJ}_{ar,bs}=-\textrm{Tr}(Y_{bs}\left[T^{IJ},Y_{ar}\right]).
\end{equation}
\indent We then consider the embedding tensor. The general form of
the embedding tensor consists of the couplings between
$\mc{A}\mc{B}$ and $\mc{B}\mc{B}$ parts. Consistency conditions
require the missing of the $\mc{B}\mc{B}$ part. Therefore, the final
form of the embedding tensor takes the form
\begin{equation}
\Theta =g\Theta_{\mc{AB}}\, .
\end{equation}
This form is similar to the embedding tensor of the same gauge group
in the $N=16$ theory studied in \cite{nicolai3}. This is not
unexpected since the $N=8$ theory considered here can be obtained
from a truncation of the $N=16$ theory. With the T-sensor given in
the appendix, we can compute the scalar potential and set up the BPS
equations from supersymmetry transformations of fermions as in the
$N=4$ theory of the previous section.
\\
\indent The full scalar manifold is 64 dimensional, but, in this
work, we are interested in the domain wall solution which involves
only the dilaton. This scalar is invariant under the $SO(8)$ part of
the gauge group. Under $SO(8)\subset (SO(8)^{(1)}\times
SO(8)^{(2)})_{\textrm{diag}}$, the 64 scalars transform as
\begin{equation}
\mathbf{8}\times \mathbf{8}=\mathbf{1}+\mathbf{28}+\mathbf{35}\, .
\end{equation}
The singlet corresponds to the dilation we are considering and
corresponds to the non-compact generator
\begin{equation}
Y_s=Y_{11}+Y_{22}+Y_{33}+Y_{44}+Y_{55}+Y_{66}+Y_{77}+Y_{88}\, .
\end{equation}
The coset representative is parametrized by
\begin{equation}
L=e^{\phi(r)Y_s}\, .
\end{equation}
As expected, the potential turns out to be
\begin{equation}
V=-192g^2e^{4\phi}
\end{equation}
which again does not admit any critical points.
\\
\indent We then find the associated domain wall solution. The metric
ansatz is again given by \eqref{DW_ansatz}. Together with the
projection condition $\gamma_r\epsilon^I=\epsilon^I$, the BPS
equation coming from $\delta\chi^{iI}=0$ gives
\begin{equation}
\phi'-2ge^{2\phi}=0\, .
\end{equation}
The solution is readily obtained
\begin{equation}
\phi=-\frac{1}{2}\ln \left(C_3-4gr\right).\label{phi_sol}
\end{equation}
Equation $\delta\psi^I_\mu=0$ reads
\begin{equation}
A'+16ge^{2\phi}=0\, .
\end{equation}
After substituting $\phi$ from \eqref{phi_sol}, we find
\begin{equation}
A=4\ln \left(C_3-4gr\right).
\end{equation}
From this, the metric takes the form
\begin{equation}
ds^2=(2gr)^8dx^2_{1,1}+dr^2\, .
\end{equation}
As in the $N=4$ case, the 8 Killing spinors are given by
$\epsilon^I=e^{\frac{A}{2}}\epsilon^I_0$ in which
$\gamma_r\epsilon^I_0=\epsilon^I_0$.
\\
\indent Similar to the $N=4$ case, after rescaling the coordinates
$x^0$, $x^1$ and $r$ to $x^0g$, $x^1g$ and $rg$ as well as changing
to the new coordinate $\rho$, we end up with the metric in the
warped $AdS_3$ form
\begin{equation}
ds^2=\rho^{-\frac{2}{3}}\ell^2
\left(\frac{dx^2_{1,1}+d\rho^2}{\rho^2}\right)
\end{equation}
where
\begin{equation}
\rho=-\frac{1}{2^6r^3}\qquad \textrm{and}\qquad
\ell^2=\frac{1}{2^{16}g^2}\, .
\end{equation}
\indent It is also interesting to study the scalar potential in more
details. We can use the truncation introduced by \cite{warner} from
which the consistency of our truncation to the $SO(8)$ singlet
follows. For example, we can consider residual symmetries of the
form $SO(4)_\textrm{diag}$, $SO(4)_\textrm{diag}\times
SO(4)_\textrm{diag}$ or $SO(3)_\textrm{diag}$. Among the 64 scalars,
there are four singlets under the $SO(4)_{\textrm{diag}}\subset
(SO(4)^{(1)+}\times SO(4)^{(1)-}\times SO(4)^{(2)+}\times
SO(4)^{(2)-})_{\textrm{diag}}$. The $SO(4)^{(1)\pm,(2)\pm}$ are
subgroups of $SO(8)^{(1),(2)}$. If we consider only $SO(3)\subset
SO(4)$ in which $\mathbf{4}\rightarrow \mathbf{3}+\mathbf{1}$, there
are eight singlets under $SO(3)_{\textrm{diag}}\subset
(SO(3)^{(1)+}\times SO(3)^{(1)-}\times SO(3)^{(2)+}\times
SO(3)^{(2)-})_{\textrm{diag}}$. Under $SO(4)^{(1)+}\times
SO(4)^{(2)+}$, there are 16 singlets. It turns out that even with
only 4 singlets, the potential computation is very complicated and
takes a very long time. So, we postpone this analysis to future
works.
\section{Conclusions}\label{conclusion}
In this paper, we have studied two domain wall solutions in
Chern-Simons three dimensional gauged supergravity. The solutions
are half supersymmetric and should correspond to dual QFT's in two
dimensions with $(4,0)$ and $(8,0)$ supersymmetries, respectively.
\\
\indent In the $N=4$ theory, the solution can be uplifted to six
dimensions via the $SU(2)$ reduction ansatz given in \cite{PopeSU2}.
And, the resulting solution in six dimensions can be thought of as a
solution of the ten dimensional heterotic string theory compactified
on $K3$ \cite{sen}. Furthermore, the six dimensional solution is
interpreted as a negative mass self-dual string according to the
discussion in \cite{PopeSU2}, see also \cite{Pope_singular_DW}.
According to the analysis on the scalar submanifolds of
$\mathbf{5}$, $\mathbf{3}_A$, $\mathbf{3}_V$ scalars, the scalar
potential, most probably, may not have any critical points at all.
Therefore, the theory may not be very useful in the study of
holographic RG flows. However, further investigations on the bigger
scalar submanifolds or even on the full scalar manifold are needed.
\\
\indent In the $N=8$ theory, after uplifting to ten dimensions, the
solution should correspond to the near horizon limit of the D1-brane
in type I theory in which the associated isometry is $ISO(1,1)\times
SO(8)$. The solution can be uplifted to ten dimensions by using the
$S^7$ reduction ansatz given in \cite{Pope_sphere} after integrating
out the 28 translational scalars corresponding to the
$\mathbf{T}^{28}$ generators. The resulting ten dimensional metric
can be used to study the dual QFT in a similar way as
\cite{Henning_AdS3S7}. Regarding the $N=8$ theory considered here as
a truncation of the $SO(8)\ltimes \mathbf{T}^{28}$ maximal gauged
$N=16$ theory via the embedding of the $SO(8,8)/SO(8)\times SO(8)$
coset into $E_{8(8)}/SO(16)$, the solution given in this work should
have the analogue in the maximal gauged supergravity. The latter
would fit into a recent classification of domain walls in maximal
gauged supergravities \cite{eric_susyDW}. Furthermore, the solutions
found in this work may provide a toy model for the study of
DW/Cosmology. Finally, the results of this paper hopefully might
give a clarification to the vacuum structure of the two gauged
supergravities.
\\
\indent In the holographic RG flow context, it is interesting to
investigate the scalar potential of the $N=8$ theory in more details
to see whether, in the absent of the dilaton, there exist any
interesting critical points and possibilities of RG flow solutions
interpolating between them. The corresponding solution should
describe a deformation of the dual $(8,0)$ two dimensional field
theory. Although this seems to be less likely at least from the
partial analysis by the present author, it would be interesting to
have a definite conclusion. Apart from the RG flow solution in the
compact gauge group $SO(4)\times SO(4)$ of \cite{bs}, no other
solutions in the $N=8$ theory are known. The $N=8$ theory with
$SO(8,8)/SO(8)\times SO(8)$ scalar manifold and non-semisimple gauge
group $(SO(4)\ltimes \mathbf{T}^6)\times (SO(4)\ltimes
\mathbf{T}^6)$ has been studied previously in \cite{gkn}, but in
that work, there have not been any possible RG flows. The flow in
this case would describe a deformation of the large $N=(4,4)$ CFT
arising from the near horizon limit of the double D1-D5 system.
\\
{\large{\textbf{Acknowledgement}}} \\
This work is partially supported by Thailand Center of Excellence in
Physics through the ThEP/CU/2-RE3/11 project and Chulalongkorn
University through Ratchadapisek Sompote Endowment Fund.
\appendix
\section{Useful formulae for $N=8$ theory}\label{detail}
In this appendix, we give some useful formulae relevant for the
construction studied in the main text. The $SO(8)$ gamma matrices
are explicitly given by
\begin{eqnarray}
\Gamma_1 &=&\sigma_4\otimes \sigma_4\otimes \sigma_4,\qquad \Gamma_2
=\sigma_1\otimes \sigma_3\otimes \sigma_4,\nonumber \\
\Gamma_3 &=&\sigma_4\otimes \sigma_1\otimes \sigma_3,\qquad \Gamma_4
=\sigma_3\otimes \sigma_4\otimes \sigma_1,\nonumber \\
\Gamma_5 &=&\sigma_1\otimes \sigma_2\otimes \sigma_4,\qquad \Gamma_6
=\sigma_4\otimes \sigma_1\otimes \sigma_2,\nonumber \\
\Gamma_7 &=&\sigma_2\otimes \sigma_4\otimes \sigma_1,\qquad \Gamma_8
=\sigma_1\otimes \sigma_1\otimes \sigma_1
\end{eqnarray}
where
\begin{eqnarray}
\sigma_1&=&\left(
             \begin{array}{cc}
               1 & 0 \\
               0 & 1 \\
             \end{array}
           \right),\qquad
\sigma_2=\left(
             \begin{array}{cc}
               0 & 1 \\
               1 & 0 \\
             \end{array}
           \right),\nonumber \\
\sigma_3&=&\left(
             \begin{array}{cc}
               1 & 0 \\
               0 & -1 \\
             \end{array}
           \right),\qquad
\sigma_4=\left(
             \begin{array}{cc}
               0 & 1 \\
               -1 & 0 \\
             \end{array}
           \right).
\end{eqnarray}
\indent The $\mc{V}$ maps are given by
\begin{eqnarray}
\mc{V}_{\mc{A}}^{ab,IJ}&=&-\frac{1}{2}\textrm{Tr}(L^{-1}J^{ab}T^{IJ}),\qquad
\mc{V}_{\mc{B}}^{ab,IJ}=-\frac{1}{2}\textrm{Tr}(L^{-1}t^{ab}T^{IJ}),\nonumber
\\
\mc{V}_{\mc{A}}^{ab,cr}&=&\frac{1}{2}\textrm{Tr}(L^{-1}J^{ab}Y^{cr}),\qquad
\mc{V}_{\mc{B}}^{ab,cr}=\frac{1}{2}\textrm{Tr}(L^{-1}t^{ab}Y^{cr}).
\end{eqnarray}
With the embedding tensor given in section \ref{N8theory}, the
T-tensor can be computed by
\begin{eqnarray}
T^{IJ,KL}&=&g\left(\mc{V}_{\mc{A}}^{ab,IJ}\mc{V}_{\mc{B}}^{ab,KL}+\mc{V}_{\mc{B}}^{ab,IJ}\mc{V}_{\mc{A}}^{ab,KL}\right),\nonumber
\\
T^{IJ,cr}&=&g\left(\mc{V}_{\mc{A}}^{ab,IJ}\mc{V}_{\mc{B}}^{ab,cr}+\mc{V}_{\mc{B}}^{ab,IJ}\mc{V}_{\mc{A}}^{ab,cr}\right)
\end{eqnarray}
where the summation over $a,b$ indices is understood.

\end{document}